# POWER SYSTEM PROBLEMS IN TEACHING CONTROL THEORY ON SIMULINK

Maddumage Karunaratne and Christopher Gabany

Department of Electrical Engineering, University of Pittsburgh, Johnstown, USA

*ABSTRACT*

*This experiment demonstrates to engineering students that control system and power system theory are not orthogonal, but highly interrelated. It introduces a real-world power system problem to enhance time domain State Space Modelling (SSM) skills of students. It also shows how power quality is affected with real-world scenarios. Power system was modeled in State Space by following its circuit topology in a bottom-up fashion. At two different time instances of the power generator sinusoidal wave, the transmission line was switched on. Fourier transform was used to analyze resulting line currents. It validated the harmonic components, as expected, from power system theory. Students understood the effects of switching transients at various times on supply voltage sinusoid within control theory and learned time domain analysis. They were surveyed to gauge their perception of the project. Results from a before/after assessment analyzed usingT-Tests showed a statistically significant enhanced learning in SSM.*

*KEYWORDS*

*Control engineering, Fourier transforms,Simulink, Power quality and Transients, Power system simulation.*

## 1. INTRODUCTION

As more and more autonomous or semi-autonomous devices, such as smart toys, pet and human robots, commercial drones, and the alike proliferate, the demand for engineers in related areas increases. Design of nearly all agile devices requires some knowledge of control methods theory. Gaining knowledge in underlying control method theories and exposure to real examples benefits undergraduates in engineering disciplines. Such skills help students effectively function as contributing members of multi-disciplinary teams, or launch their own entrepreneur careers.

All the senior Electrical Engineering (EE) students and some Computer Engineering (CE) seniors take one semester course on Control Systems. The associated laboratory class utilizes MATLAB control tool boxes such as symbolic manipulations and Laplace functions. For modeling in State Space (SS) and differential equation solving, students use Simulink in MATLAB. Being a tool, MATLAB/Simulink is explained and explored within lab experiments only. Students generally show interest when real-world problems are presented and solved.

One such real problem is the transient overvoltages caused by switching operations of transmission lines which are of fundamental importance in selecting equipment insulation levels and surge-protection devices for power engineers. Therefore, the understanding of the nature of transmission line transients, and analysis of power quality, are important [1, 2]. A relatively straight forward problem in power systems is transient overvoltages resulting from switching a transmission or distribution line that has largely capacitive characteristics, common with systems featuring cable (which is inherently capacitive) and/or capacitor banks (for voltage support, power factor correction, and/or power flow management). Switching may induce a power quality event. It is important to ensure that high quality power is being delivered to the customer. A single power quality event may cause the affected industries up to millions of monetary loss [3].

Students with prior understanding of circuits and signals benefit from the new experiment by solving and seeing the technical details about the behavior of power supply lines under switching conditions. The need to regulate and filter power supply voltages would be appreciated when they are being incorporated with sensitive low voltage system components. Switching in electric power transmission systems occur for a multitude of reasons, some planned (e.g. maintenance) and others unplanned (e.g. faults, such as a trees or animals contacting energized components initiating a short circuit that is subsequently cleared by a protective device). Regardless of the cause, de-energization and re-energization (restoration of electrical power via closing a circuit breaker or switch) could introduce significant switching events. These events cause electrical transients, which may produce dangerous high frequency overvoltages. The transient voltage and frequency shifts will be a function of the transmission cable capacitance, source inductance, source voltage, and the remaining system parameters.

The magnitude of the transient depends on the precise instant (relative to zero crossing point of sinusoidal waveform of the power generator) when the circuit breaker closes, turning on the power. The maximum effect occurs when power is switched on at a peak voltage in its sinusoidal waveform. Switching transient analysis determines the risk of the magnitude of the transient overvoltages, and the time duration of the high frequency transient which usually is in the range of 100 Hz to 10,000 Hz [1]. The nominal power supply frequency is 60 Hz. Since a control methods course analyzes transient response of various systems [4], this experiment introducing a real world scenario benefits both majors.

Power system engineers employ several commercial software tools for electrical system simulations, streamlining power system studies and transient events. Existing literature provides many examples of using dedicated power system software packages such as Power System Computer Aided Design (PSCAD) [5], used in this paper, for transients and power quality investigations.

However, power system transient analysis using Simulink has been very limited, and the analysis typically uses SimscapePowerSystems$^{TM}$ blocks in Simulink. One such paper [6] explains simulations of various power quality events on power systems, utilizing Simulink SimPowerSystems, which was later renamed SimscapePowerSystems (SPS). This SPS provides electrical component libraries and analysis functions for modeling and simulating electrical power systems, enabling rapid creation of models of physical systems within the Simulink environment [7]. While literature exists [8, 9] on control theory course experiments, none was found to apply power system problems into control theory experiments.

This lab project presents an alternative method to solve and analyze power system switching transient overvoltages using the general Simulink tool without using the dedicated extension of Simulink for power systems (SPS). Then, the results are compared against the results obtained from another software tool (PSCAD) which is commercially dedicated to analyze power systems.This hybrid power/control system experiment revealed that the transient studies produce time domain based results, regardless of underlying computational techniques. This experiment provided the students a platform to practice this skill using Simulink time domain analysis and SSM, although most of their learning experience has been in frequency domain analysis - Laplace transform.

Advantages of dedicated power system analysis tools, such as PSCAD mentioned in this paper, (used only by the authors to validate this experiment), and SPS, include providing the user a more graphical user-friendly interface, but their 'engine', the software algorithms, are built with the same analytical theorems used in this project. This work reveals to students, that no matter the modern power system analysis software used, the analysis is fundamentally rooted in the same circuit analysis attributes.

The Student Outcomes of this experiment include:

(i) Learn the effects of power system switching events, extending the understanding of power quality when transients occur, beyond the exposure to Laplace transforms. Appreciate the application of Fourier analysis on real-world problems.
(ii) Learn that Simulink has the capability to provide solutions similar to solutions provided by dedicated power systems simulations software. They are both mathematical models with certain approximations.
(iii) Further understand the meaning of natural frequency and damping, as applied to a real-world system.
(iv) Practice SSM, by following the topology of a system, without using a set of abstract differential equations.

While Laplace domain would have sufficed to model this power system problem, the SS model was developed and emphasized to show its flexibility in following a circuit topology. Also, this multi-stepped experiment avoids that simplest Laplace route to achieve several Student Outcomes as stated in objectives. One outcome is to practice SS representation of an electrical system using a bottom up approach (part 2 of the experiment walks the student through the network components and models the signals (currents and voltages), using only the $1^{st}$ order derivatives as is the form of SS in control theory). As indicated by survey responses, the assignment also reinforces the meaning of natural frequency of a system, as in $2^{nd}$ order systems, by way of Laplace transform methods. Using a multiple input multiple output (MIMO) system might have demonstrated the ultimate value of Simulink in SS modeling. However, attempting to model and analyze a MIMO system in Laplace could have exceeded the time and scope of the course topics.

To measure if this lab project would enhance students' learning of the State Space concepts, the authors administered a closed book pencil/paper quiz one day before the lab experiment. The assessment was on an electrical network yielding a second order SS representation. Students were not given any feedback on that quiz, nor were they aware that it would be given again. The same assessment was administered the day after the lab experiment. Both quizzes were graded, and individual scores were analyzed statistically to correlate two sets of scores. It is noted that, no student has taken the power systems course at the time of this experiment.

Section 2 of the paper briefly describes the content of the course on major categories of control systems theory and the expected course outcomes for undergraduate engineering students.Section 3 elaborates the experiment conducted, in order to model the power system, along with subsequent Fourier Transform analysis, highlighting the adverse frequency components in the line current,which degrade the power quality. The solutions and results are in Section 4 of the paper. Section 5 ponders the results of a student survey at the end of the experiment, to understand and interpret how the control class students viewed this (power system related) experiment. Results from the statistical analysis on the enhancement of SSM skills are presented in Section 6 while Section 7 states concluding remarks with some thoughts for future work.

## 2. COURSE CONTENT AND OUTCOMES

The course exposes undergraduates to Control System theory and provides a basic understanding of its applications. The emphasis is on modeling and analysis of systems rather than pure design aspects due to the nature of the program. The students' exposure appears sufficient to work in the industry where motors, pumps, boilers and other controlled large equipment are commissioned or maintained. The course consists of the following sections.

*Section A:* Frequency domain modeling using Laplace transforms to develop transfer functions for electrical networks (amplifiers, too), and different types of mechanical systems.
*Section B:* Model linear and nonlinear systems in time domain as SS representations in Simulink. Conversion among transfer functions, differential equations, and SS techniques are practiced.

*Section C:* Time response and transient analysis on $2^{nd}$ order systems, along with impacts of poles and zeros.

*Section D:* Reduction of subsystems to analyze overall behavior. Feedback systems are analyzed and designed to achieve required closed loop characteristics.

*Section E:* Stability criteria and techniques are taught, including Routh-Hurwitz techniques.

*Section F:* Output response error analysis is studied to determine steady state and disturbance errors. Students learn how to design a controller to achieve desired transient responses for

Section C and to eliminate or reduce steady state errors for various input function types.

## 3. THE MODELLING EXPERIMENT

Projects and experiments typically reflect on the material covered in the lecture class. Students use Simulink for some of the experiments. The experiment was conducted after students had studied and explored the topics of SS representation of second and higher order systems in lectures and in other assignments. Presented below is a brief summary of the new experiment encompassing control and power system theory, titled Simulink modeling of a transient current in a power system being switched on. The goals of the assignment are:

(a) Understand and compare the second order system behaviors and parameters under real world example conditions and parameters.
(b) Model in SS representation form of an electrical power system circuit by directly mapping its circuit topology instead of a derived (abstract) differential equation.
(c) Observe adverse impacts of transients on the power quality in the transmission line when the power source is switched on, after a power failure due to an accident or manually.
(d) Subject the transient current to Fourier analysis to obtain its frequency harmonics to explore the impact on the power quality.

### 3.1. Problem Statement

A 3-phase 345kV rated 60 Hz electrical substation is connected to a 45-mile-long transmission cable supplying power to a remote substation, which steps down the voltage for local distribution. To simplify, a single phase schematic representation is provided in Figure 1. Cable capacitance and substation loads are lumped and marked by capacitance C, and resistance R2. The transmission cable introduces a small inductance L2 and small resistance R1 to the system. The power generator has inherent inductance represented by L1 on supply side. A switch represents the actual closing of a circuit breaker to re-energize the cable and substation. The natural frequency of the circuit is excited by the switching operations, with energy oscillating between inductance and capacitance. However, with resistances R1 and R2 present in the circuit, the natural response fades away leaving only the forced response of 60Hz utility power, Vg(t). Figure 1 shows voltages marked at circuit nodes: Vg at generator, Vs at switch, V1 between lumped L2 and lumped R1, and voltage Vc at substation. The cable current, i(t) starts after the switch closure.

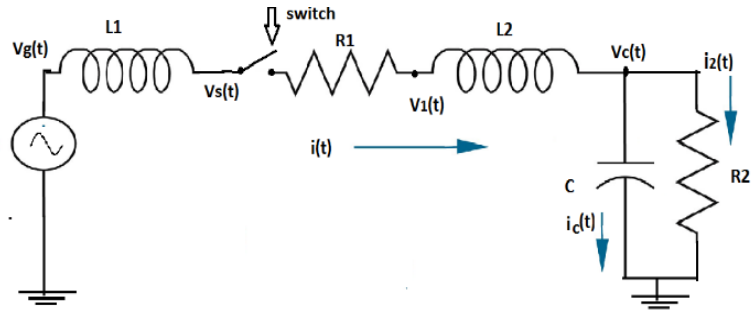

Figure 1. Transmission Line with Network Parameters

### 3.2. Project Assignment

*Part 1:* Using the parameters L1, L2, R1, R2, and C, obtain the Laplace transfer function for I(s)/Vg(s). Find its natural frequency, Wn using the fact that cable resistance is very small compared to end customer load resistances, i.e. R2 is much larger than R1. Also calculate Wn using L1=25 mH, L2=0.1 mH, R1= 0.5 Ω, R2=5000 Ω, and C = 24 μF.

*Part 2:* SS representation is based on 1st order derivatives. The power on switch is modelled using a product block with a step function. Obtain Vs(t) using the 60Hz sine source, Vg(t) and line current, i(t) in L1. Amplitude of Vg(t) is 345*√(2/3) kV. To obtain di/dt in SS form, derive voltage drop V1(t) in L2 as Vs(t) – R1.i(t). Subtracting Vc(t) from V1(t) gives the voltage across L2 to obtain di/dt. Vc(t) is proportional to integral of ic(t), where ic(t) = i(t) – i2(t). R2 derives i2(t) under Vc(t) completing the model. Collecting data using LineCurOut (to Simulink Workspace) i(t) is needed for Fourier analysis. To reduce data volume, turn the switch on after 15 complete cycles of Vg(t), at t=0.25 seconds. However, to switch it on at Vg(t) peak, a step function is turned on another ¼ cycle later at t=0.25 + 1/240 seconds.

*Part 3:* When the switch is closed at peak input, observe the collapsing of Vs(t) and the subsequent overshoot. Using i(t) waveform, measure peak to peak times to calculate the oscillatory transient frequency. Observe the i(t) transients die down after a few seconds as expected from a $2^{nd}$ order underdamped system. Compare the frequency of damped oscillations on i(t) waveform with the derived value in part 1. This is the natural frequency, Wn of the utility power system.

*Part 4:* Use a MATLAB script to obtain the Fourier analysis of the line current by running a fast Fourier transform (FFT). Through Simulink Workspace, save a sufficient number of data points between 0 ~ 0.5 sec time scale for FFT. Obtain a graph of the magnitude of the frequency components and note the dominant and secondary frequencies. Compare the natural frequency from part 1 with this secondary frequency.

*Part 5:* Rerun part 2 model after adjusting the switching on time to be at a 0 crossing (0.25 sec) and obtain a plot of i(t). Rerun FFT script to obtain final frequency components and show that the instance of reenergizing the line impacts the power quality such as frequency shifts, and overvoltages.

### 4. RESULTS OF THE EXPERIMENT

Figure 2 shows the State Space model developed in Simulink by students. Input sine waveform source is configured to have a sampling time of 0.0001s to obtain sufficient data samples. Figure 3 depicts the supply voltage waveform at the switch, Vs(t), for a short period of time, which has become stable after 15 cycles. The switching event occurs after another ¼ cycles later at peak voltage of Vg(t). It clearly reveals the instant the transmission cable was reenergized, and the resulting instantaneous voltage collapse followed by the overshoot to almost twice the peak value of the stable Vg(t). The high frequency oscillatory transient overvoltage remains, 'riding' on the primary 60 Hz frequency. As the simulation time increases, approximately 10 cycles after the switching event, the transient overvoltage decays (due the damping presence of network resistance), and the primary 60 Hz waveform again begins to take shape.

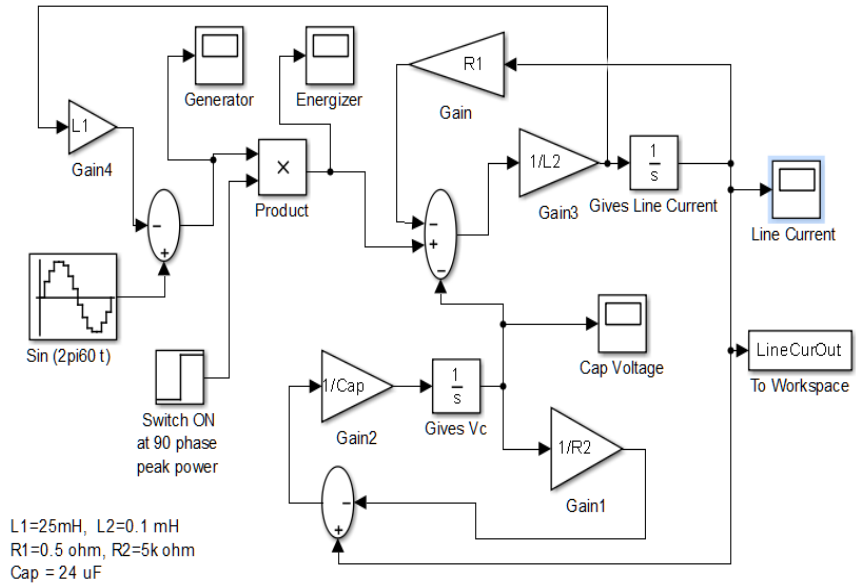

Figure 2. Simulink model of the system

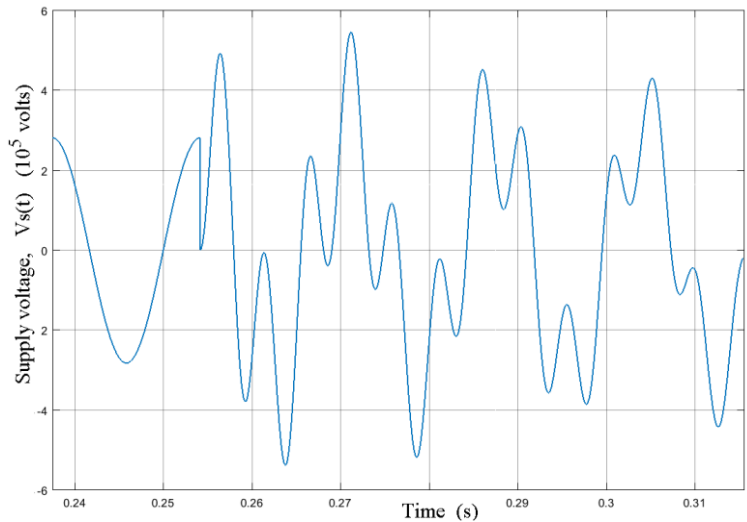

Figure 3. Supply voltage waveform at switching node showing
the instanteneouscollapse of the voltage at t = 0.2542s.

Figure 4shows the line current for a very short time interval. It was used for part 3 of the experiment. The distortions on the 60 Hz input sine wave appears to have been modulated by a higher frequency signal. The graph indicates an 11,000A maximum current peak at 0.265s time and a distorted signal with highest frequency of 200Hz calculated by taking 8 (peak to peak) cycles as marked on Figure 4.

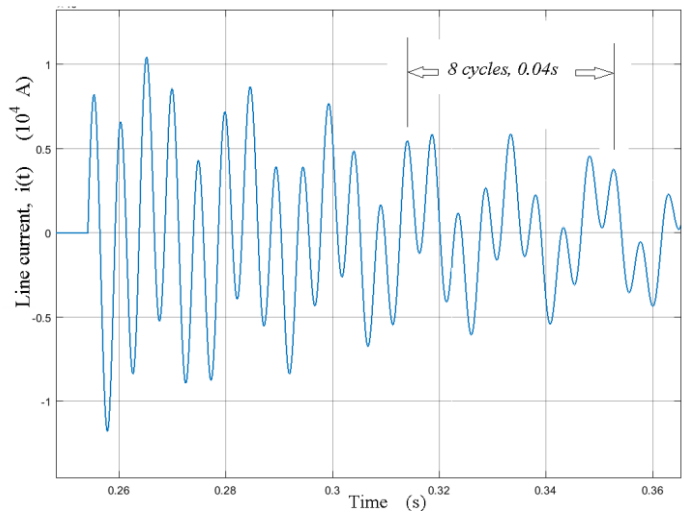

Figure 4. Line current waveform showing switching transients

Authors validated the accuracy of the Simulink model using the academic version of the industry grade PSCAD power systems modeling tool, and the results are presented in Figure 6. Top graph shows the line current, i(t) and bottom shows the supply voltage at the switch, Vs(t). Both Simulink and PSCAD models used the same timing event to energize the cable line at peak voltage after 15¼ cycles of starting the simulation. As in Simulink, it clearly reveals the instant the transmission cable was energized, and the resulting instantaneous collapse of Vs(t) followed by the overshoot to almost twice the peak value. Evidently and as expected, Simulink results match PSCAD precisely.

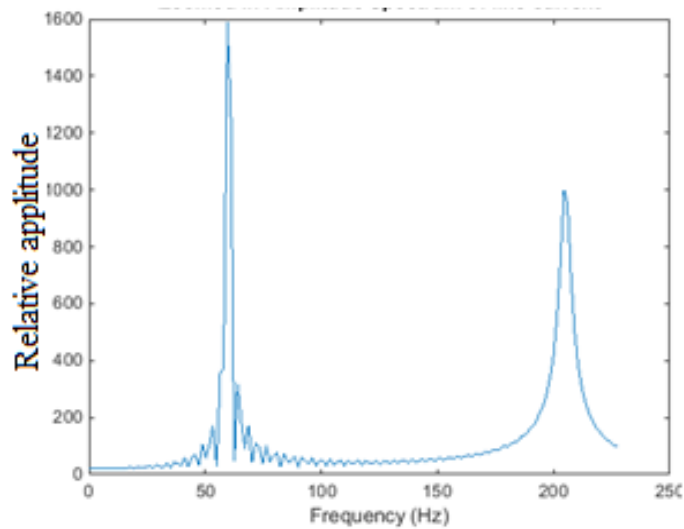

Figure 5. Amplitude of line current frequency components

It is worth noting that high frequency transient switching overvoltage signatures (both frequency and peak overvoltage) are largely determined by the system capacitance [10]. In this model, a transmission line cable was energized (cables inherently have a high capacitance), and it may be assumed the distribution substation featured power factor correction capacitors. Both sources of capacitance were modelled as a lumped parameter, C (at the substation). The bottom graph shows the line current which agrees with the Simulink results on all the aspects such as overcurrent values, modulating transient frequency (of 205 Hz), and final steady state current of (2500 amp) validating the Simulink model and the techniques used in here.

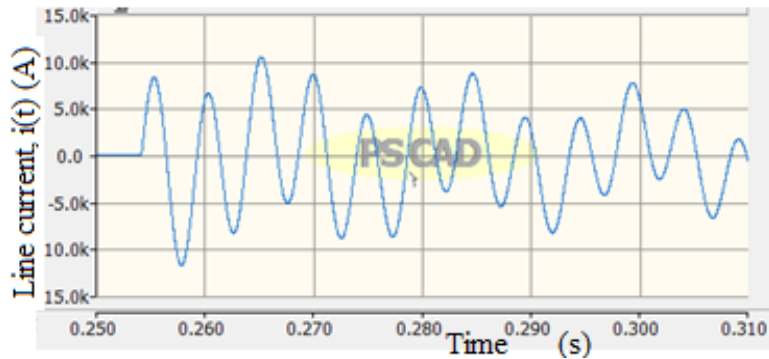

Figure 6 (a) Line current waveform from PSCAD

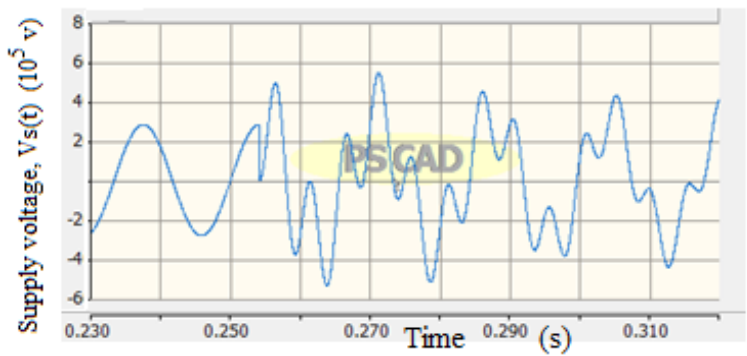

Figure 6 (b) Supply voltage waveform from PSCAD

Figure 7 shows the line current for a very short time interval obtained from step 5 of the lab experiment. The power source is switched on when its sinusoidal signal value is at 0 volts (or almost zero) for this case while Figure 2 shows the case that signal is at the peak. The waveform exhibits much less distortion than Figure 4 with reduced peaks, and current settles to steady state value of about 2500 Amps much faster. Overcurrent reaches only 5000A as opposed to 11,000A in Figure 2. Also Figure 7 graph provides 10 full cycles of the modulating signal between 0.25 sec and 0.3s time interval providing again close to 205Hz higher harmonic frequency.

This comparison demonstrates that the worst voltage fluctuation and the highest current with unwarranted frequencies may occur depending on when a utility line is re-energized. This is a real-world scenario. For a three-phase switching event, with three phases each displaced by 120 electrical degrees, the exact time instant that event occurs is largely uncontrollable although modern 'single-pole switching' circuit breakers are available [10]. Fourier analysis of line current samples showed the same frequency components as in Figure 5, but with much less amplitude for secondary frequency.

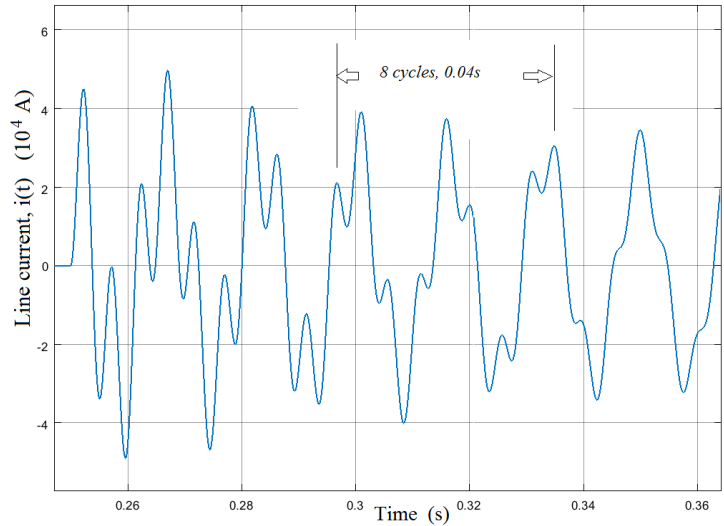

Figure 7. Line current transients with less distortion

## 5. STUDENT FEEDBACK ON THE EXPERIMENT

Students were surveyed at the end of the lab project to gauge the effectiveness of the experiment, and how they perceived modeling an electrical power system, within the context of control system concepts, using a more general tool (Simulink) than dedicated power system tools. Twenty-four students participated in this experiment. About half were electrical and the rest were computer engineering majors. No attempt was made to differentiate results based on their majors, since their course assessment data does not show a correlation to the major. The results are summarized qualitatively in Table I. The responses were weighted based on the typical rubrics of 1 to 5 (1-strongly disagree, 2-disagree, 3-neutral, 4-agree, 5-strongly agree). Column 2 of Table I lists what each question was trying to assess from students' perception and knowledge. The 3rd column shows the average rubric score given by students for each question.

Table 1. Summary of student survey results

|    | The Essence of the question asked | Ave |
|----|-----------------------------------|-----|
| 1  | It helped me further understand the behavior of $2^{nd}$ order systems | 4.1 |
| 2  | I learned what the natural frequency means. | 4.2 |
| 3  | Now I have a better understanding of transient (or switching) oscillations in a $2^{nd}$ order system. | 3.9 |
| 4  | This project made clearer the idea of the *natural response* and *forced response* of a system. | 4.0 |
| 5  | I now appreciate Fourier analysis more. | 3.9 |
| 6  | This Simulink modelling work made it clearer what state space representation is. | 3.9 |
| 7  | I understood more how to model in State Space. | 3.7 |
| 8  | I learned about power quality and how it can be affected by switching transients. | 3.7 |
| 9  | I already knew switching transients on power lines (internship/other courses) | 2.5 |
| 10 | I have a better appreciation of plugged in electronic devices since they have to withstand such frequency shifts occasionally. | 4.2 |
| 11 | I am surprised that Simulink can give same results as dedicated power system analysis tools. | 4.1 |
| 12 | I gained a better appreciation of control systems theory after seen that it is present in normal utility power systems – not just in contrived assignments. | 4.4 |

## 6. LEARNING ENHANCEMENT DATA

Both quizzes were graded and normalized to a 100% grading scale that was divided into 5 brackets of 20% spread per bracket. The horizontal axis in Figure 8 shows the 5 brackets along with the % of students in each 20% grade spread (from aprior assessment). The left (*Before*) vertical bar in each bracket provides the 1st quiz average of students in each bracket. The right (*After*) bar shows the 2nd quiz average of the students (originally placed in the bracket).

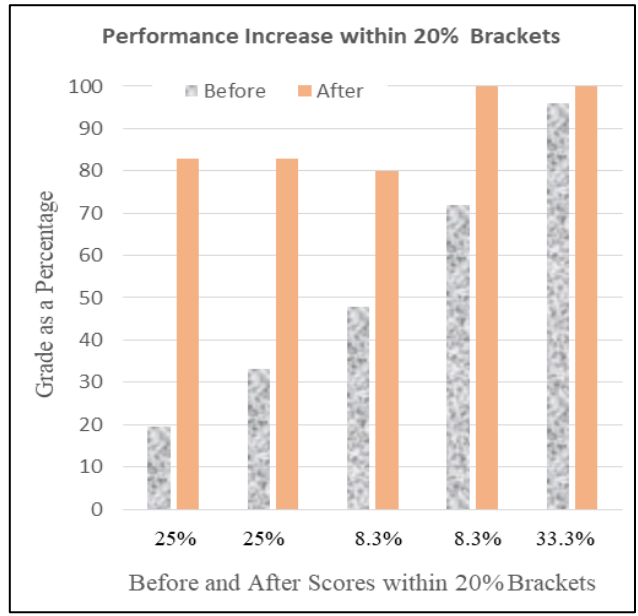

Figure 8. Spread of average quiz scores per bracket

The two sets of averages in the 5 brackets shown in Figure 8 were analyzed using Two-Sample T-Test in MiniTab$^{TM}$ software. However, the data in 3rd and 4th brackets were combined for this T-Test analysis due to their low number of data points (only 8.3% in each).Table II provides results of analysis of the performance data on 24 students who also took the survey listed in Table I.The first row of Table II is thebracket number related to Figure 8. The 2nd and 3rd rows provide the mean values of scores in brackets, before and after the experiment, respectively. The next two rows show the standard deviations of the data corresponding to 2nd and 3rd rows. The next two rows provide T-Value and P-Value parameters of T-Test results, respectively. The very low P-Value in each bracket asserts that the learning enhancement achieved from this control theory on power system experiment is significant at above 95% confidence level.

Table 2. Statistical T-Test Results

| Bracket | 1 | 2 | 3 &4total | 5 |
|---|---|---|---|---|
| 1st Mean | 18.7 | 33.3 | 60 | 96 |
| 2nd Mean | 82.6 | 82.7 | 90.9 | 100 |
| 1stStDev | 2.1 | 5.5 | 13.8 | 4.2 |
| 2ndStDev | 12.6 | 20.7 | 10.3 | 0.01 |
| T-Value | -12.3 | -5.67 | -3.59 | -2.71 |
| P-Value | 0.000 | 0.002 | 0.016 | 0.030 |

## 7. CONCLUSION

This project measured its effectiveness in two ways: (i) A student survey after the experiment, and (ii) a direct assessment (before and after) of learning. In the survey, students agreed they learned, from the experiment, more Control Methods subject (response to questions 1~4) and signal theory (question 4). They feel that they are still challenged in learning SS representation based on responses to questions 6 and 7. The answer to question 9 shows that only a very few students were aware of power system switching transients before this experiment. Responses to questions 11 and 12 were encouraging since the survey show students were not indifferent to this experiment, and they were pleasantly surprised.

The assessments done prior to the lab showed a low level of students' performance on SSM skills. After the experiment, student assessment data revealed a significant level of enhanced skills, among all the five brackets that were based on ranges of students' first set of scores. Figure 8 indicates a failing level of performance, as well as excellent levels for the 1$^{st}$ quiz. The experiment boosted all students to very good and excellent levels. Even those 33% of students in the 5$^{th}$ bracket showed a statistically significant improvement due to the experiment.

The power system transient study produced time domain based results, regardless of underlying computational techniques. This experiment provided the students a platform to practice this skill using Simulink time domain analysis (SS) while most of their prior experiences were in frequency domain – Laplace. Advantages of dedicated power system analysis tools such as PSCAD mentioned in this paper (not used by students, but by the authors), provide the user a more graphical user-friendly interface, but their 'engines' (software algorithms) are built within the same analytical theorems used in this project.  This experiment reveals to students, that no matter the path of the Electrical or Computer Engineer post-secondary education, whether control systems, power systems, electronics, etc., the disciplines are fundamentally rooted in the same circuit analysis attributes.

Overall, this experiment was viewed as a success by students, and direct assessments also indicate the same. Therefore, this laboratory project would continue in the future control system course offerings. As future work, authors plan to include a Simscape PowerSystems based experiment in the Power Systems II follow up course, and present the experience in a suitable venue.

## AUTHORS

**Maddumage Karunaratne** (M'92) received the B.S. degree in electronics & telecommunication engineering from University of Moratuwa, Sri Lanka and the M.S. degree in electrical engineering from University of Mississippi, MS. He earned the Ph.D. degree in electrical engineering from the University of Arizona, Tucson, AZ, in 1989. Before joining the faculty of the University of Pittsburgh at Johnstown, Pennsylvania, USA, he worked for several Silicon Valley, CA semiconductor companies, including Intel and Apple., as a senior engineering consultant in design and test.

**Christopher Gabany** is an Electrical Engineering professor at the University of Pittsburgh at Johnstown, Pennsylvania, USA. He received his BSEET from the same college 1994, and the MSEE from Michigan Technological University in 2013.He has a PE (Professional Engineer) license and CEM, CEA, and CPQ certificates through AEE. He is an Officer in the Army National Guard, and currently a PhD candidate at the University of Pittsburgh.